# Single and Double Diffractive Production of Dilepton and Photon at LHC

Gongming Yu[1*], Rabia Hameed[2], Liyuan Hu[2], Qiang Hu[3]

[3]College of Physics and Technology, Kunming University, Kunming 650214, China
[2]Fundamental Science on Nuclear Safety and Simulation Technology Laboratory, Harbin Engineering University, Harbin 150000, China
[3]Institute of Modern Physics, Chinese Academy of Sciences, Lanzhou 730000, China
Email: ygmanan@163.com, rabiahameedofficial@gmail.com, liyuanhu91@163.com, qianghu@impcas.ac.cn

We have investigated the single and double diffractive production of dileptons and photons in ultra-peripheral collisions at the Large Hadron Collider (LHC). Utilizing advanced theoretical models that integrate quantum electrodynamics (QED) and Quantum Chromodynamics (QCD) frameworks, we analyze the differential cross sections of these processes, with particular emphasis on the role of the Pomeron and resolved Pomeron structures, as well as resolved photon structures. Our research employs diffractive production mechanisms to predict dilepton and photon production rates under various LHC energy scenarios. Our results demonstrate distinct production patterns for single and double diffractive processes, highlighting their potential as probes for studying the electromagnetic structure of heavy ions and the dynamics of soft interactions in high-energy collisions. This paper provides new insights into the photon-mediated and Pomeron-mediated production mechanisms and sets the stage for future experimental investigations at collider facilities.

## I. INTRODUCTION

In the ultra-peripheral collision process of two heavy ions, leptons are one of the fundamental particles that make up matter, including electrons, muons, and tauons and their corresponding neutrinos. Dilepton refers to a lepton pair composed of a lepton and its antiparticle.[1] These dileptons are produced in collisions at particle accelerators such as RHIC and LHC by various processes, including decaying or destroying heavier particles. A primary particle, the photon, is essential to studying high-energy physics and the interactions in particle colliders, such as the Large Hadron Collider (LHC)[2, 3]. At these high-energy facilities, photons are fundamental particles that carry the electromagnetic force, play a crucial role in several events that occur during collisions [4, 5]. Diffractive physics has been extensively explored in hadron-hadron collisions, analyzing various final states such as di-jets, electroweak vector bosons, dileptons, heavy quarks, and quarkonium paired with a photon, among others. A significant theoretical approach within this field is the Resolved Pomeron Model, initially proposed by Ingelman and Schlein[6-8]. This model supports the diffractive factorization formalism[9] and suggests that the Pomeron possesses a partonic structure, comprising quarks and gluons. In this study, we apply the Resolved Pomeron

Model for diffractive photon production[10]. This model assumes that the Pomeron possesses a distinct partonic structure, facilitating hard processes within a Pomeron-proton interaction for single diffractive events, or between two Pomerons in double diffractive scenarios [7, 8].

This research focuses on the photoproduction of dileptons and photons in ultraperipheral collisions (UPC) at high-energy physics facilities like the RHIC and LHC [11, 12]. UPCs, where heavy ions generate minimal physical contact but intense electromagnetic fields, facilitate the production of dileptons and photons through intricate electromagnetic interactions. The study employs the Resolved Pomeron Model to analyze the partonic interactions within Pomerons, crucial for understanding single and double diffractive processes in these collisions [13-16]. Dilepton photoproduction, a key Quantum Electrodynamics (QED) process, is explored by observing how virtual photons from the electromagnetic fields of proximate heavy ions materialize into real dilepton pairs [16-18]. The research method, the equivalent photon approximation, simplifies the complexities of photon emission and absorption processes, aiding in calculating the total differential scattering cross-section[19]. Further, the study incorporates Regge theory[20, 21] and QCD-based factorization to analyze and predict particle behavior under various kinematic conditions[21-23], including the motion, angular distribution, and correlations in dilepton invariant mass [24, 25].

This paper is organized as follows: In section II, we present the semi-coherently two-photon production process for dileptons at RHIC and LHC energies. The numerical results for dileptons production are plotted in Sec.III. Finally, the conclusion is given in Sec.IV.

## II. GENERAL FORMALISM

Diffractive production of dileptons in high-energy physics involves the generation of a pair of leptons in a diffractive process, such as those between protons at high-energy colliders like the Relativistic Heavy Ion Collider (RHIC) and Large Hadron Collider (LHC) [26-28]. These processes are characterized by the exchange of a Pomeron or a photon, leading to a final state with distinctive features, such as rapidity gaps and often the survival of initial state particles, such as protons, in a largely unaltered state [17]. There are three mechanisms of diffractive dilepton production:

- Pomeron-Induced Dilepton Production,
- Photon-Induced Dilepton Production, and
- Pomeron-Photon Interactions.

Pomeron-Induced Dilepton Production involves the exchange of a colorless, gluonic entity, which can interact with quarks within a proton to produce dileptons through quark-antiquark annihilation or gluon fusion processes [29]. Photon-Induced Dilepton Production occurs through the interaction of quasi-real photons emitted by colliding protons. Here is the flowchart Table 1, that first defined for dilepton and then for Photon. As long diffractive processes are concerned, in hadron collisions they are well described, with respect to the overall cross sections, by Regge theory in terms of the exchange of a Pomeron with vacuum quantum numbers[30].

The Pomeron-nucleon coupling vertex is derived from the fundamental theory of strong interaction[31], QCD. Pomeron-Nucleon Coupling Vertex is a function representing the interaction between a Pomeron ($\mathbb{P}$) and a nucleon (N, typically a proton or neutron). In Regge theory, the Pomeron is a trajectory rather than a particle, representing a family of exchanged particles that dominate at high energies[8, 32-34]. The Pomeron flux factor, the parton distribution functions for both proton and nucleus, and the Pomeron trajectory will all contribute to determining the probability and kinematics of the process. The Mandelstam variables are fundamental in the study of particle physics, providing a mathematical framework to describe the energy and momentum transfer during particle collisions. These variables, typically denoted as *s, t,* and *u*, encapsulate the kinematics of two-to-two particle scattering processes. The variable *s* represents the square of the total energy in the center-of-mass system, while *t* and *u* correspond to the squares of the momentum transfer between the particles involved.

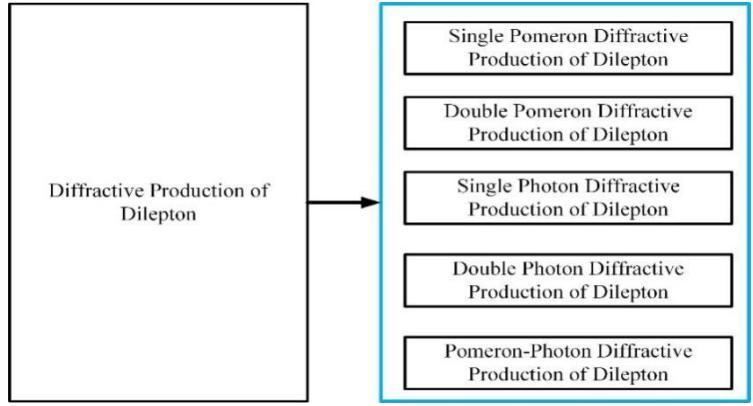

**Table 1** Flowchart for Diffractive Production of Dilepton

Where $\hat{s}$, $\hat{t}$ and $\hat{u}$ contains the equation (1), (2), and (3).

$$\hat{s} = (p_a' + p_b)^2 = p_a'^2 + 2p_a' \cdot p_b + p_b^2 = 2p_a' \cdot p_b = 2z_a\ x_b(s/2) = x_a z_a\ x_b s, \quad (1)$$

$$\hat{t} = (p_c - p_a')^2 = p_c^2 - 2p_a' \cdot p_c + p_a'^2 = M^2 - 2x_a p_c \cdot p_A = M^2 + x_a z_a(t - M^2) = M^2 - x_a z_a x\ s, \quad (2)$$

$$\hat{u} = (p_c - p_b)^2 = p_c^2 - 2p_b \cdot p_c + p_b^2 = M^2 - x_b p_c \cdot p_B = M^2 + x_b\ (u - M^2) = M^2 - x_b x_1\ s, \quad (3)$$

Putting the values of $\hat{s}$, $\hat{t}$ and $\hat{u}$, we will get the value of $M^2$.

$$M^2 = \hat{s} + \hat{t} + \hat{u}. \quad (4)$$

Starting with the form factor equation and pomeron trajectory. The proposal of a phenomenological Pomeron exchange model with a vector-type Pomeron-nucleon vertex has been made;

$$V^{P-N} = \beta \gamma^\mu F_1(\hat{t}). \quad (5)$$

Here $\beta$ is the coupling strength of the Pomeron ($\mathbb{P}$) to nucleon (*N*, typically a proton or neutron). $\gamma^\mu$ is the part of vector-type vertex, indicating the Lorentz structure of the interaction, but in quantum field theory, $\gamma^\mu$ are the Dirac matrices [35, 36], that describes fermions like nucleons. Sometimes, $\beta = 3\beta_0$ with $\beta_0$ being the coupling strength of the Pomeron-coupling quark[37, 38]. The factor 3 stems from quark numbers inside the nucleon according to quark counting rule. $F_1(\hat{t})$ is the iso-scalar nucleon form factor and is taken to be the dipole form with the values of $A_p = 2.8, B_p = 0.7, \beta = 3.24 \text{GeV}^2$. This factor is a common choice in phenomenological models for describing the spatial distribution and internal structure of the nucleon. The form factor typically decreases with increasing $(\hat{t})$ reflecting the fact that the probability of a high-momentum transfer interaction is lower due to the finite size of the nucleon.

The parameters $A_P$ and $B_P$ shape the t-dependence of the form factor $F_1(\hat{t})$. According to

$$F_1(\hat{t}) = \frac{4m_P^2 - A_P(\hat{t})}{(4m_P^2 - \hat{t})\left(1 - \frac{\hat{t}}{B_P}\right)^{-2}}, \tag{6}$$

the Regge theory, the Pomeron trajectory, $\alpha_{\mathbb{P}}(\hat{t})$, related to the propagator of Reggeon in the s-channel and to a physical particle or resonance in t-channel can be expressed as;

$$\alpha_{\mathbb{P}}(\hat{t}) = \alpha_{\mathbb{P}}(0) + \alpha'_{\mathbb{P}} t, \tag{7}$$

The trajectory $\alpha_{\mathbb{P}}(0) = 0.085$, $\alpha'_{\mathbb{P}} = 0.25 \text{GeV}^{-2}$, $\alpha_{\mathbb{P}}(\hat{t}) \simeq 1.08 + 0.25(\hat{t})$ describes the trajectory of the pomeron in the complex angular momentum plane as a function of the momentum transfer squared ($\hat{t}$). $m_p$ is representing the mass of proton, $m_p = 1.67 \times 10^{-27}$ kg. The pomeron trajectory $\alpha_{\mathbb{P}}(\hat{t})$ describes the kinematics of diffractive process[39, 40]. $f_{\mathbb{P}/p}(x, Q^2)$ is representing a standard Pomeron flux factor from Regge phenomenology, that describes the probability of a pomeron being emitted by a proton with a fraction $x$ of the proton's momentum transfer $Q^2$. This factorization scale is related to the energy scale of the interaction and affects the running of the strong coupling constant in Quantum Chromodynamics (QCD).

Parton Distribution Functions (PDFs)[41, 42], are fundamental in describing the internal structure of hadrons in terms of quarks and gluons (collectively known as partons). They are crucial for predicting the outcomes of high-energy collisions. Pomeron exchange, in high-energy physics, means the pomeron is a hypothetical object used to describe the strong force in the framework of Regge theory, particularly in processes with no quantum number exchange.

$$f_{\frac{\mathbb{P}}{p}}(x, Q^2) = A_{\mathbb{P}} \frac{e^{B_{\mathbb{P}} t}}{x^{2\alpha_{\mathbb{P}}(\hat{t})-1}}, \tag{8}$$

The parameters $A_{\mathbb{P}}$ and $B_{\mathbb{P}}$ are the Pomeron flux [43, 44]. Moreover, the parton distribution function PDF for proton is;

$$f_{\frac{\mathbb{P}}{p}}(x, Q^2) = \frac{9\beta^2}{4\pi^2}\left(\frac{1}{x}\right)^{2\alpha_{\mathbb{P}}(\hat{t})-1} [F_1(\hat{t})]^2, \tag{9}$$

where $\beta$ is the Pomeron coupling strength to the proton [45]. $\alpha_{\mathbb{P}}(\hat{t})$ shows in equation (9). The Pomeron flux factor contributes to the probability of the Pomeron emission, while the PDF accounts for the distribution of partons in the proton that participate in the hard scattering process leading to dilepton production [46]. $A_{\mathbb{P}}$ is the normalization parameter for Pomeron flux factor. The factorization scale for the strong coupling and for the evaluation of PDFs is $\mu_F^2$. And it represents; $\mu_F^2 = \sqrt{p_T + m^2}$, $p_T$ is the transverse momentum of particle, and $m$ is its mass. The PDF for a nucleus $f_{\mathbb{P}/N}(x, Q^2)$ represents the probability density of finding a parton (quark or gluon) with a fractional momentum $x$ inside the nucleus. This function depends on both the momentum fraction $x$ and the scale $Q^2$, similar to the PDF for a proton. For a Pomeron, one begins with the Regge trajectory;

$$\alpha_{\mathbb{P}}(\hat{t}) = 1 + \epsilon + \alpha'_{\mathbb{P}} t, \tag{10}$$

where $\epsilon = 0.085$, $\alpha'_{\mathbb{P}} = 0.25 \text{GeV}^{-2}$. Note that this trajectory has an intercept close to 1, suggesting that the Pomeron behaves like a spin-one boson in processes in which the exchanged four-momentum squared $k^2 = t$ is small. Pomeron range parameter is;

$$r_o^2 = \alpha'_{\mathbb{P}} \ln(s/m^2). \tag{11}$$

Following Donnachie and Landshoff, we denote the Nucleon-Pomeron coupling $\beta_{N\mathbb{P}}$ by[47, 48];

$$\beta_{N\mathbb{P}} = 3\beta_0 F_N(-t), \tag{12}$$

where $\beta_0$ is the quark-nucleon coupling and $F_N(-t)$ is the isoscalar magnetic nucleon form factor for the nucleon. For a Nucleus-Pomeron interaction, we appeal to the additivity of the total nucleon-nucleon cross sections and replace $\beta_0$ by $A\beta_0$ and replace $F_N(-t)$ by the elastic nuclear form factor $F(\vec{k}^2)$. Here $\vec{k}^2$ is the squared momentum transfer in the interaction. $Q_0^2$ is a parameter characterizing the scale of the nuclear form factor. $\beta_0$ and $F_N(-t)$ for nucleus-Pomeron interactions accounts for the fact that a nucleus is composed of multiple nucleons[49, 50]. The nuclear form factor describes how the nucleus as a whole responds to the Pomeron exchange and the associated momentum transfer.

$$F(\vec{k}^2) = e^{-\vec{k}^2/2Q_0^2}, \tag{13}$$

The parton distribution function for nucleus with mass $M$ and nucleon number $A$ is:

$$f_{\frac{\mathbb{P}}{N}}(x, Q^2) = \left(\frac{3A\beta_0 Q_o^2}{2\pi}\right)\left(\frac{\hat{s}}{m_p^2}\right)\frac{1}{x}e^{-x^2 M_N^2/Q_0^2}, \tag{14}$$

$Q_0$ represents the pomeron-nucleon coupling constant. $\hat{s}$ denotes the invariant sub-process with which the pomeron participates. $\beta_0$ is the quark-nucleon coupling. Putting $Q_0 = 60.0\text{GeV}$, $\beta_0 = 1.8\text{GeV}^{-1}$. The exponential term $e^{-x^2 M_N^2/Q_0^2}$ accounts for the spatial distribution of partons within the nucleus. Now, we have to explain partonic structure of pomeron and parton distribution function of the nucleus, which represent the probability of finding a quark or a gluon with a fraction $x$ of the pomeron's momentum[51]. The pomeron is a hypothetical particle that mediates the strong interaction between hadrons. The partonic structure of the pomeron is given by two functions;

$$f_{\frac{q}{\mathbb{P}}}(x) \text{ and } f_{\frac{g}{\mathbb{P}}}(x),$$

The quark distribution function is given by:

$$f_{\frac{q}{\mathbb{P}}}(x) = A_1 x^{A_2}(1-x)^{A_3}, \tag{15}$$

Here the parameters for this function are $A_1 = 0.066$, $A_2 = 0.29$, and $A_3 = 0.72$. The gluon distribution function is given by:

$$f_{\frac{g}{\mathbb{P}}}(x) = B_1 x^{B_2}(1-x)^{B_3}, \tag{16}$$

Where the parameters for this function are $B_1 = 1.22$, $B_2 = 3.13$ and, $B_3 = 0.31$. These values are obtained from a theoretical model based on perturbative Quantum Chromodynamics (pQCD) and factorization theory. The parton distribution function of the nucleus, $f_N(x)$, describes the probability of finding a parton (quark or gluon) with a fraction of the nucleus's momentum. The correction factor $R(x, A)$ modifies the parton distribution to account for nuclear effects, such as shadowing and nuclear binding. Also, parton distribution function for the nucleus $f_N(x)$ combines the parton distributions of protons and neutrons within the nucleus, weighted by their relative numbers.

$$f_N(x) = R(x, A)\left[\frac{Z}{A}f_P(x) + \frac{N}{A}f_n(x)\right], \tag{17}$$

Where $Z$ and $N$ are the numbers of protons and neutrons in the nucleus, and $f_P(x)$ and $f_n(x)$ are the parton distribution functions of the proton and neutron, respectively. The function $R(x, A)$ contain:

$$R(x,A) = 1 + 1.19 ln^{\frac{1}{6}} A[x^3 - 1.5(x_o + x_L)x^3 + 3x_o x_L x] - \left[\alpha_A \frac{1.08\left(A^{\frac{1}{3}} - 1\right)}{ln(A+1)} \sqrt{x}\right] e^{-x^2/x_0^2}, \quad (18)$$

Where $x_o = 0.1$, $x_L = 0.7$, $\alpha_A = 0.1\left(A^{\frac{1}{3}} - 1\right)$. The parton distribution function of the nucleon:

$$xf(x) = A_0 x^{A_1}(1-x)^{A_2}, \quad (19)$$

Here, $x$ is the momentum fraction, and $A_0$, $A_1$, and $A_2$ are parameters that are determined by fitting the equation to experimental data. The function $f(x)$ gives the probability density of finding a parton with momentum fraction $x$ inside the proton. The equation shows the parton distribution function for different types of partons (*d, u, s,* and *g*) in the nucleon.

Table 2 Parameters used for PDF of nucleon.

| | | | |
|---|---|---|---|
| $d_v$ | $A_o = 1.4473$ | $A_1 = 0.6160$ | $A_2 = 4.9670$ |
| $u_v$ | $A_O = 1.7199$ | $A_1 = 0.5526$ | $A_2 = 2.9009$ |
| $\bar{u} + \bar{d}$ | $A_o = 0.0616$ | $A_1 = -0.2990$ | $A_2 = 7.7170$ |
| $\bar{d}/\bar{u}$ | $A_o = 33657.8$ | $A_1 = 4.2767$ | $A_2 = 14.8586$ |
| $s = \bar{s}$ | $A_o = 0.0123$ | $A_1 = -0.2990$ | $A_2 = 7.7170$ |
| $g$ | $A_o = 30.4571$ | $A_1 = 0.5100$ | $A_2 = 2.3823$ |

### Single-Pomeron Diffractive Production of Dilepton

In Single Diffractive Processes, a collision occurs between two particles *A* and *B,* where one particle remains mostly intact, losing only a small amount of momentum and energy, while the other may break apart or produce new particles. This type of interaction is facilitated by the exchange of a theoretical entity known as a Pomeron ($\mathbb{P}$), which is key in understanding high-energy scattering events in particle physics. While dileptons are particle pairs consisting of two leptons (such as electrons or muons). The term "dilepton" indicates that the focus of this study is on the production processes involving pairs of leptons. To derive the main equation for the single pomeron diffractive production process for dileptons, we will start with the main equation (20).

$$\frac{d\sigma_{AB \to l^+l^-X}^{Sin.Pom.Diff}}{dM^2} = \int dx_a \, dz_a \, dx_b d\hat{t} f_{\underset{A}{\mathbb{P}}}(x_a, Q^2) f_{\underset{\mathbb{P}}{a'}}(z_a, \mu_F^2) f_{\underset{B}{b}}(x_b, Q^2) \frac{d\hat{\sigma}}{dM^2 d\hat{t}}(a'b \to l^+l^-X), \quad (20)$$

The given expression represents a differential cross-section for the single Pomeron diffractive production of a lepton pair ($l^+l^-$) in a collision between two particles (*A* and *B*). The distribution functions $f_{\mathbb{P}/A}(x_a, Q^2)$, $f_{a'/\mathbb{P}}(z_a, \mu_F^2)$, and $f_{b/B}(x_b, Q^2)$ representing the probability of finding respective partons with given momentum fractions and scales.

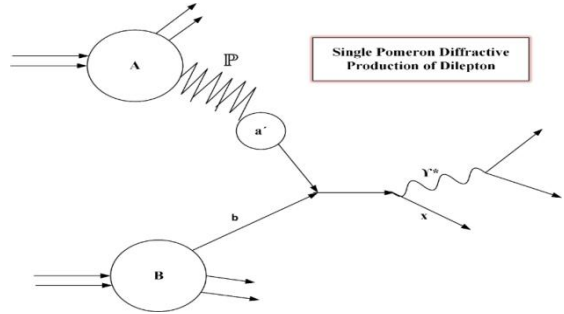

**Figure 1** Single Pomeron Diffractive Production of Dilepton ($l^+l^-$).

Here $f_{\mathbb{P}/A}(x_a, Q^2)$, is a parton distribution function (PDF) for the Pomeron ($\mathbb{P}$) in hadron $A$, depending on the momentum fraction $x_a$ and the scale $Q^2$, usually related to the energy scale of the interaction, mathematically represented by equation (9) and (14). In Figure 1 Single Pomeron Diffractive Production of Dilepton, two particles, $A$ and $B$, interact via a pomeron exchange, which is a process that occurs without significantly altering the structure of the interacting particles. The pomeron exchange leads to the creation of a system labeled as $a'$ and $b$, with $a'$ going on to interact further to produce a virtual photon (denoted by $\gamma^*$), that decays into a lepton and its antiparticle ($l^+l^-$), conserving charge, energy, and momentum. $f_{a'/\mathbb{P}}(z_a, \mu_F^2)$, is the PDF for a parton $a'$ in the Pomeron, depending on $z_a$ and the factorization scale $\mu_F^2$, denoted by equation (15) and (16). And $f_{b/B}(x_b, Q^2)$ is the PDF for a parton in hadron $B$, mathematically can be represented by equation (9) and (14). Here, $z_a$ is used to denote the fraction of the pomeron's momentum that is carried by a parton (denoted as $a'$) involved in the hard scattering process

The term $d\hat{\sigma}/dM^2 d\hat{t}$ represents the hard scattering cross-section for the process. $d\hat{\sigma}/dM^2 d\hat{t}(a'b \to l^+l^-X)$ represents the partonic level differential cross-section for the production of the dileptons from partons. $d\hat{\sigma}$ denotes the differential cross-section for the specific parton-parton interaction. The cross-section is a measure of the probability of a certain interaction occurring. In this case, it's the probability of partons $a'$ and $b$ producing dileptons and possibly other products. $a'b \to l^+l^-X$ notation describes the reaction under consideration. Partons $a'$ and $b$ are the initial particles involved in the collision, and $l^+l^-X$ represents the final state, including a lepton ($l^+$), its antiparticle ($l^-$), and other possible products ($X$). $d\hat{t}$ represents the differential of a Mandelstam variable $\hat{t}$. Starting with the sub-process cross sections $d\hat{\sigma}/d\hat{t}$ and then incorporate the provided kinematic variables. Equation (21) represents the differential cross section for the process where a quark and an antiquark ($q\bar{q}$) annihilate to produce a virtual photon (denoted as $\gamma^*$) and a gluon. The factors $\alpha_s$ involve the strong coupling constant the electromagnetic coupling constant ($\alpha$), the charge $e_f^2$ squared of the quark flavor and the Mandelstam variables.

$$\frac{d\hat{\sigma}}{d\hat{t}} q\bar{q} \to \gamma^* g = \frac{8}{9}\frac{\pi\alpha\alpha_s e_f^2}{\hat{s}^2}\left(\frac{\hat{t}}{\hat{u}} + \frac{\hat{u}}{\hat{t}} + \frac{2M^2\hat{s}}{\hat{t}\hat{u}}\right), \tag{21}$$

Equation (22) is for the process where a quark and an antiquark ($q\bar{q}$) annihilate to produce a virtual photon and a real photon ($\gamma^*\gamma$). The main difference here is the replacement of the strong coupling constant with the electromagnetic coupling constant, reflecting the electromagnetic nature of the interaction.

$$\frac{d\hat{\sigma}}{d\hat{t}} q\bar{q} \to \gamma^*\gamma = \frac{2}{3}\frac{\pi\alpha^2 e_f^2}{\hat{s}^2}\left(\frac{\hat{t}}{\hat{u}} + \frac{\hat{u}}{\hat{t}} + \frac{2M^2\hat{s}}{\hat{t}\hat{u}}\right), \tag{22}$$

Equation (23) represents the differential cross section for the process where a quark scatters off a gluon to produce a virtual photon and another quark. The structure of the equation is similar to the equation (21), but with different ratios of the Mandelstam variables.

$$\frac{d\hat{\sigma}}{d\hat{t}} qg \to \gamma^* q = \frac{1}{3}\frac{\pi\alpha\alpha_s e_f^2}{\hat{s}^2}\left(-\frac{\hat{u}}{\hat{s}} - \frac{\hat{s}}{\hat{u}} - \frac{2M^2\hat{t}}{\hat{s}\hat{u}}\right) \tag{23}$$

In order to derive the main equation of single diffractive production of dilepton we need to define the variables $x_1$ and $x_2$. That are clearly shown in equation (24) and (25).

$$x_1 = \frac{M^2 - u}{s} = \frac{1}{2}\left(x_T^2 + 4\tau\right)^{\frac{1}{2}} e^y, \tag{24}$$

$$x_2 = \frac{M^2-t}{s} = \frac{1}{2}\left(x_T^2 + 4\tau\right)^{\frac{1}{2}} e^{-y}, \tag{25}$$

Putting the values of $x_T$ and $\tau$ that are explain in equation (26), we can get variables $x_1$ and $x_2$.

$$x_T = \frac{2p_T}{\sqrt{s}}, \tau = \frac{M^2}{s}. \tag{26}$$

Equation (24) and (25) becomes:

$$x_1 = \frac{M^2-u}{s} = \frac{1}{2}\left(\left(\frac{2p_T}{\sqrt{s}}\right)^2 + 4\left(\frac{M^2}{s}\right)\right)^{\frac{1}{2}} e^{y}, \tag{27}$$

$$x_2 = \frac{M^2-t}{s} = \frac{1}{2}\left(\left(\frac{2p_T}{\sqrt{s}}\right)^2 + 4\left(\frac{M^2}{s}\right)\right)^{\frac{1}{2}} e^{-y}, \tag{28}$$

where $p_T$ is the transverse momentum $M$ is the invariant mass of dilepton. $y$ is the rapidity. By using the values of equation (27) and (28), we will get parton's momentum fraction $x_a, x_b$ and $z_a$.

$$x_a = \frac{x_b x_1 - \tau}{x_b z_a - x_2 z_a}, \tag{29}$$

$$x_b = \frac{x_a z_a x_2 - \tau}{x_a z_a - x_1}, \tag{30}$$

$$z_a = \frac{x_b x_1 - \tau}{x_b x_a - x_2 x_a}. \tag{31}$$

Energy-momentum relation for a dilepton final state is:

$$E^2 = p_T^2 + p_L^2 + M^2. \tag{32}$$

Here $p_T$ and $p_L$ are the transverse and longitudinal momentum components, respectively, and M is the invariant mass shown in equation (4). Jacobi determinant, which is a mathematical tool used to transform the integration variables in a multidimensional integral. In this case, it relates the variables $z_a, \hat{t}, \theta$ and $x_T^{'}$ and it is used to calculate the differential cross-section for the production of dilepton. The equation (33) shows that the differential cross-section depends on the momentum fractions $x_a, x_b, z_a$ and $x_2$, the center-of-mass energy s, the transverse momentum $x_T^{'}$, and the scattering angle $\theta$.

$$dz_a d\hat{t}\left|\frac{D(z_a,\hat{t})}{D(\theta,x_T^{'})}\right| d\theta dx^{'} = \begin{vmatrix} \frac{\partial z_a}{\partial \theta} & \frac{\partial z_a}{\partial x_T^{'}} \\ \frac{\partial \hat{t}}{\partial \theta} & \frac{\partial \hat{t}}{\partial x_T^{'}} \end{vmatrix} = \frac{x_a z_a x_b}{x_a x_b - x_a x_2} \cdot \frac{s}{2} \cdot \frac{x_T^{'}}{\sin\theta} d\theta dx_T^{'}, \tag{33}$$

$x_a, z_a,$ and $x_b$ are represented mathematically by equation (29), (30), and (31). The value of dilepton $l^+ l^-$ are given in equation (34).

$$l^+ l^-: \frac{d^3p}{E} = \frac{dp_x dp_y dp_z}{E} = d^2p_T dy = \pi dp_T^2 dy = \pi \frac{s}{2} \frac{x_T^{'}}{\sin\theta} d\theta dx_T^{'}, \tag{34}$$

The longitudinal momentum $p_z$ is determined by the mass $M$ of the dilepton, its transverse momentum $p_T$, and its rapidity.

$$p_z = \left(M^2 + p_T^2\right)^{\frac{1}{2}} \sinh y. \tag{35}$$

Total energy $p_o$, as shown in equation (36) is determined by the mass, transverse momentum $p_T$, and rapidity $y$ of the dilepton.

$$p_o = (M^2 + p_T^2)^{\frac{1}{2}} \cosh y \Rightarrow p_z = p_o \tanh y = y \; p_o = y \, E \Rightarrow dp_z = E dy. \quad (36)$$

Single pomeron diffractive production of dilepton in nucleus-nucleus collisions. These equations are related to the calculation of the cross-section for the single pomeron diffractive production of dilepton in nucleus-nucleus collisions.

$$\frac{d\sigma_{AB \to l^+l^-X}^{Sin.Pom.Diff}}{dM^2 d^2 p_T dy} = \int dx_a \, dx_b f_{\underline{\mathbb{P}}}(x_a, Q^2) f_{\underline{a}'}(z_a, \mu_F^2) f_{b/B}(x_b, Q^2) \frac{x_a z_a x_b}{x_a x_b - x_a x_2} \times \frac{d\hat{\sigma}}{dM^2 d\hat{t}}(a'b \to l^+l^-X), \quad (37)$$

$$\frac{d\sigma_{AB \to l^+l^-X}^{Sin.Pom.Diff}}{dM^2 d^2 p_T dy} = \int dx_a \, dx_b f_{\underline{\mathbb{P}}}(x_a, Q^2) f_{\underline{a}'}(z_a, \mu_F^2) f_{b/B}(x_b, Q^2) \frac{x_a z_a x_b}{x_a x_b - x_a x_2} \times \frac{\alpha}{3\pi M^2} \sqrt{1 - \frac{4m_l^2}{M^2}} (1 + \frac{2m_l^2}{M^2}) \frac{d\hat{\sigma}}{d\hat{t}}(a'b \to \gamma^* X), \quad (38)$$

**Double-Pomeron Diffractive Production for Dilepton**

The differential cross-section for the double pomeron diffractive production process of dileptons in the collision between particles *A* and *B*, as a function of the squared invariant mass $M^2$ of the final state. Variables $x_a$ and $x_b$, are the momentum fractions of partons in particles *A* and *B*, and $z_a$ and $z_b$ are the momentum fractions of partons within the proton. In Figure 2 Double Pomeron Diffractive Production of Dilepton, two hadrons labeled *A* and *B* interact without directly colliding, by exchanging two pomerons ($\mathbb{P}$). These pomerons, representing colorless exchanges typically associated with gluons in Quantum Chromodynamics (QCD), lead to a diffractive event where the hadrons remain intact (*a'* and *b'*). This interaction creates an intermediate state *X*, which then decays into a virtual photon ($\gamma^*$), that finally produces a dilepton pair ($l^+ \, l^-$). Mathematically, can be written as:

$$d\sigma_{AB \to l^+l^-X}^{Dou.Pom.Diff}/dM^2.$$

$f_{\mathbb{P}/B}(x_b, Q^2), f_{\mathbb{P}/A}(x_b, Q^2)$, are Parton Distribution Function (PDF) for partons in particle *A* and *B* within the proton shown in equation (8) and (14), as a function of momentum fraction $x_a$ and $x_b$ with the factorization scale $Q^2$. $f_{a/\mathbb{P}}(z_a, \mu_F^2)$, $f_{b/\mathbb{P}}(z_a, \mu_F^2)$, shown in equation (15) and (16) are PDF for partons in the proton as a function of momentum fraction $z_a$ and $z_b$ with the factorization scale $\mu_F^2$. ($ab \to l^+l^-X$), representing the differential partonic cross-section for the production of dileptons. While putting the values of these Mandelstam variables we can easily find $M^2$. Jacobi determinant defined in equation (33). The value of dilepton $l^+ \, l^-$ are given in equation (34). Total energy $p_o$, as shown in equation (36) while The longitudinal momentum $p_z$ in equation (35). $d\hat{t}$ variable represents the squared four-momentum transfer between the incoming and outgoing particles.

$$\frac{d\sigma_{AB \to l^+l^-X}^{Dou.Pom.Diff}}{dM^2} = \int dx_a \, dz_a dx_b dz_b d\hat{t} f_{\underline{\mathbb{P}}}(x_a, Q^2) f_{\underline{a}}(z_a, \mu_F^2) f_{\underline{\mathbb{P}}}(x_b, Q^2) f_{\underline{b}}(z_b, \mu_F^2) \times \frac{d\hat{\sigma}}{dM^2 d\hat{t}}(ab \to l^+l^-X), \quad (39)$$

The sub-process cross sections $d\hat{\sigma}/d\hat{t}$ are given in equation (21), (22), and (23). Equation (39) denotes production of dilepton through double pomeron process. $l^+ \, l^-$ denotes production of dilepton, we have in equation (34). $\hat{s}, \hat{t},$ and $\hat{u}$ are the Mandelstam variables for the partonic collision (before parton fragmentation), contain in equation (1), (2), and (3). Here $x_1, x_2$ are the momentum fractions of the partons in the pomeron involved in the collision shows in (27) and (28). $z_a$ and $z_b$ represents the momentum fractions of the partons in the proton. Where $p_T$ is the transverse momentum *M* is the invariant mass of dilepton. *y* is the rapidity.

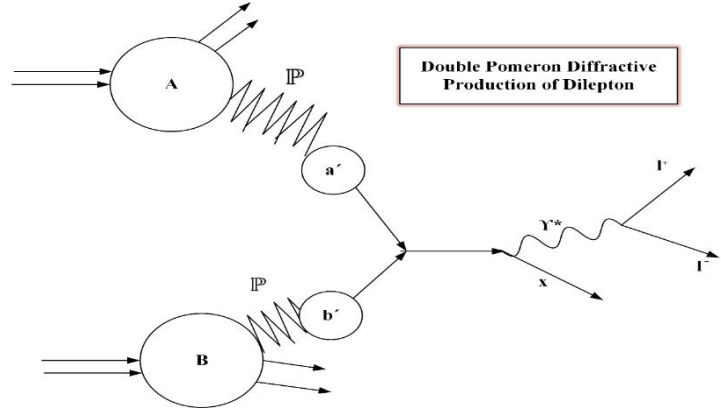

**Figure 2** Double Pomeron Diffractive Production of Dilepton ($l^+l^-$).

By using the values of $x_1$, $x_2$ from equations (27) and (28), we will get parton's momentum fraction $x_a, x_b, z_a$ and $z_b$.

$$x_a = \frac{x_b z_b x_1 - \tau}{x_b z_b z_a - x_2 z_a}, \tag{40}$$

$$x_b = \frac{x_a z_a x_2 - \tau}{x_a z_a z_b - x_2 z_b}, \tag{41}$$

$$z_a = \frac{x_b z_b x_1 - \tau}{x_a z_b x_b - x_2 x_a}, \tag{42}$$

$$z_b = \frac{x_a x_1 z_a - \tau}{x_b x_a z_a - x_2 x_b}. \tag{43}$$

By using all the equations, we can easily get the production of dilepton $l^+\ l^-$ through double pomeron process. The formula (44) represents the cross-section for the production of dileptons (electron-positron pairs) in double diffractive nucleus-nucleus collisions. The cross-section for the dilepton produced by the double diffractive in nucleus-nucleus collisions can be written as:

$$\frac{d\sigma_{AB \to l^+l^-X}^{Dou.Pom.Diff}}{dM^2} = \int dx_a\, dx_b dz_b f_{\frac{\mathbb{P}}{A}}(x_a, Q^2) f_{\frac{a'}{\mathbb{P}}}(z_a, \mu_F^2) f_{\frac{\mathbb{P}}{B}}(x_b, \mu_F^2) f_{\frac{b'}{\mathbb{P}}}(z_b, \mu_F^2) \frac{x_a z_a x_b z_b}{x_a x_b z_b - x_a x_2} \times \frac{\alpha}{3\pi M^2} \sqrt{1 - \frac{4m_l^2}{M^2}} (1 + \frac{2m_l^2}{M^2}) \frac{d\hat{\sigma}}{d\hat{t}}(a'b' \to \gamma^*X). \tag{44}$$

## Single-Photon Diffractive Production for Dilepton

The process begins with a diffractive interaction, which is characterized by the exchange of a Pomeron ($\mathbb{P}$) or a photon between the colliding particles shown in Figure 3. This interaction is such that the colliding particles remain intact or only slightly excited, and there is a rapidity gap—a region in the detector with no detected particles—indicating the non-destructive nature of the interaction. In single-photon production, a photon is emitted from one of the protons. This photon can then interact with the electromagnetic field of the other proton or with a virtual photon emitted by the other proton, leading to the production of a dilepton pair through the process.

The provided equations describe the differential cross section for single-photon diffractive production of dileptons in high-energy collisions. The expression involves parton distribution functions (PDFs), such as $f_{a'/\gamma}(x)$ and $f_{\gamma/p}(x)$, kinematic factors, and relations defining parton momentum fractions. The equations incorporate constants, masses, and parameters relevant to the theoretical framework. Parton distribution functions (PDFs), such as $f_{a'/\gamma}(x)$ and $f_{\gamma/p}(x)$ describe the probability of finding a particular parton (like a quark, gluon, or photon) inside a hadron (like a proton, neutron, or nucleus) with a given fraction of the hadron's momentum. The variable $z_a$ represents the fraction of the photon's momentum carried by the parton $a'$ and $x$ represents the fraction of the proton's momentum carried by the photon. $\mu_F$ is the factorization scale, a parameter in quantum field theory calculations. The parton distribution function of parton is represented by:

$$f_{\frac{a'}{\gamma}}(z_a, \mu_F^2).$$

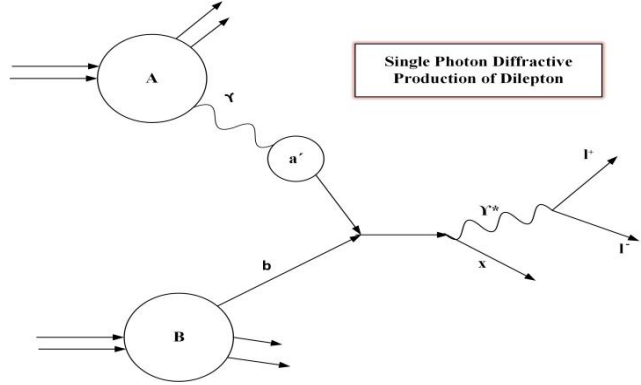

Figure 3 Single Photon Diffractive Production of Dilepton ($l^+l^-$).

Equation (45) gives the distribution of photons within a proton. The expression involves the fine-structure constant, the momentum fraction $x$ and a logarithmic term dependent on $Q_{min}^2$, which relates to the minimum momentum transfer.

$$f_{\frac{\gamma}{p}}(x) = \frac{\alpha}{2\pi}\left(1 + \frac{(1-x)^2}{x}\right)\left(\ln A_p - \frac{11}{6} + \frac{3}{A_p} - \frac{2}{2A_p^2} + \frac{1}{32A_p^3}\right), \tag{45}$$

Here $Q_{min}^2$ is minimum momentum transfer squared, a key quantity in determining the lower limit of the momentum transfer in the interaction. $Q_{min}^2$ is given in equation (46), $A_p$ is shown in equation (47).

$$Q_{min}^2 = -m_p^2 + \frac{1}{2s}[(s+m_p^2)(s-xs+m_p^2) - (s-m_p^2)\sqrt{(s-xs+m_p^2)^2 - 4m_p^2 xs}] \sim \frac{m_p^2 x^2}{1-x}, \tag{46}$$

The value of $A_p$ is shown in equation (47).

$$A_p = 1 + \frac{0.71 GeV^2}{Q_{min}^2}, \tag{47}$$

By putting the values of $Q_{min}^2$, $A_p$. We can get PDF of photon in proton $f_{\gamma/p}(x)$. Similar to the photon distribution in a proton, but for a nucleus $f_{\gamma/N}(w)$. It depends on the nuclear charge Z, and involves logarithmic terms with $w, \gamma_L, R$. $\gamma_L$ is a factor related to the Lorentz boost of the nucleus which are defined in terms of the nucleon-nucleon center-of-mass energy $\sqrt{s_{NN}}$, proton mass $m_p$ and nuclear radius $R$ often approximated as $R = 1.2 \times A^{1/3}$ fm where $A$ is the mass number of the nucleus. $w$ is the fraction of the nucleon-nucleon center-of-mass energy carried by the photon in the nuclear case. By putting the values from equation (49), we will get (48).

$$f_{\frac{\gamma}{N}}(w) = \frac{2Z^2\alpha}{\pi w} \ln\left(\frac{\gamma_L}{wR}\right), \tag{48}$$

While we have;

$$\gamma_L = \frac{\sqrt{S_{NN}}}{2m_p}, \quad w = x\frac{\sqrt{S_{NN}}}{2}, \quad R = 1.2A^{\frac{1}{3}}, \tag{49}$$

Here $x_1, x_2$ are the momentum fractions of the partons in the pomeron involved in the collision shows in (27) and (28). Using these we can get kinematics variables $x_a, x_b, z_a$.

$$x_a = \frac{x_b x_1 - \tau}{x_b z_a - x_2 z_a}, \tag{50}$$

$$x_b = \frac{x_a z_a x_2 - \tau}{x_a z_a - x_1}, \tag{51}$$

$$z_a = \frac{x_b x_1 - \tau}{x_b x_a - x_2 x_a}, \tag{52}$$

Using all the terms, we will get an equation (53) for single photon diffractive production of dilepton.

$$\frac{d\sigma_{AB \to l^+l^-X}^{Sin.Pho.Diff}}{dM^2 dp_T^2 dy} = \int dx_a \, dx_b f_{\frac{\gamma}{A}}(x_a, Q^2) f_{\frac{a'}{\gamma}}(z_a, \mu_F^2) f_{\frac{b}{B}}(x_b, Q^2) \frac{x_a z_a x_b}{x_a x_b - x_a x_2} \frac{d\hat{\sigma}}{dM^2 d\hat{t}}(a'b \to l^+l^-X), \tag{53}$$

$$\frac{d\sigma_{AB \to l^+l^-X}^{Sin.Pho.Diff}}{dM^2 dp_T^2 dy} = \int dx_a \, dx_b f_{\frac{\gamma}{A}}(x_a, Q^2) f_{\frac{a'}{\gamma}}(z_a, \mu_F^2) f_{\frac{b}{B}}(x_b, Q^2) \frac{x_a z_a x_b}{x_a x_b - x_a x_2} \times \frac{\alpha}{3\pi M^2} \sqrt{1 - \frac{4m_l^2}{M^2}} (1 + \frac{2m_l^2}{M^2}) \frac{d\hat{\sigma}}{d\hat{t}}(a'b \to \gamma^* X), \tag{54}$$

## Double-Photon (photon-photon) Diffractive Production for Dilepton

In this process, two photons, emitted from the protons, interact with each other to produce a pair of leptons in a diffractive manner, with the protons typically remaining intact or only slightly excited. In high-energy pp collisions, protons can emit photons due to the intense electromagnetic fields present during their approach. These photons are quasi-real, meaning they have very low virtuality, behaving almost like real photons. The emitted photons can interact with each other, leading to the production of dilepton pairs as shown in Figure 4. This is a quantum electrodynamics (QED) process, governed by the electromagnetic interaction. The initial proton-proton interaction is diffractive because it involves the exchange of a colorless, neutral entity (in this case, photons) leading to a final state characterized by rapidity gaps — regions in the detector where no particles are produced. The protons themselves remain largely unaffected by the interaction, either staying intact or being excited into a state that does not break them apart. The expression (59) describes the double-photon (photon-photon) diffractive production for dileptons in the context of the differential cross section.

The expression $AB \to l^+l^-X$ represents the differential cross section for the process in double-photon diffractive production. By putting all the values of variables $x_a, z_a, x_b, z_b$, represented in equation (55) to (58), along with various parton distribution functions (PDFs) we will get the main equation of our focus. These relations describe the momentum fractions in terms of the partonic subprocesses and the squared invariant mass $M^2$. The expression includes a factorized product of PDFs, parton momenta fractions, and an alpha ($\alpha$) factor.

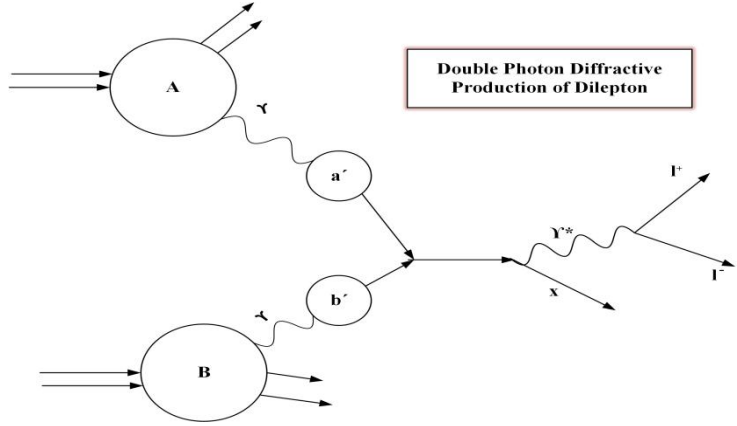

**Figure 4** Double Photon Diffractive Production of Dilepton ($l^+l^-$).

$f_{\gamma/A}(x_a, Q^2)$, $f_{\gamma/B}(x_b, Q^2)$, are the parton distribution functions for photons in the colliding particles $A$ and $B$ shown in equation (45) and (48). $f_{a'/\gamma}(z_a, Q^2)$, $f_{b'/\gamma}(z_b, Q^2)$, are the parton distribution functions for partons in the photon $a'b'$, respectively. $M^2$ is the invariant mass of the dilepton system, $p_T$ is the transverse momentum, and $y$ is the rapidity. $Q^2$ represents the virtuality scale, and $m_l$ is the mass of the charged lepton.

$$x_a = \frac{x_b x_1 z_b - \tau}{x_b z_a z_b - x_2 z_a}, \tag{55}$$

$$x_b = \frac{x_a z_a x_2 - \tau}{x_a z_a z_b - x_2 z_b}, \tag{56}$$

$$z_a = \frac{x_b x_1 z_b - \tau}{x_b x_a z_b - x_2 x_a}, \tag{57}$$

$$z_b = \frac{x_a x_1 z_a - \tau}{x_b x_a z_a - x_2 x_b}. \tag{58}$$

Using these kinematics variables and PDF's along with cross-section process, we will get the desire equation of double diffractive production of dilepton as shown in equation (59).

$$\frac{d\sigma_{AB \to l^+l^-X}^{Dou.Pho.Diff}}{dM^2 dp_T^2 dy} = \int dx_a dx_b dz_b f_{\frac{\gamma}{A}}(x_a, Q^2) f_{\frac{a'}{\gamma}}(z_a, Q^2) f_{\frac{\gamma}{B}}(x_b, Q^2) f_{\frac{b'}{\gamma}}(z_b, Q^2)$$

$$\frac{x_a z_a x_b z_b}{x_a x_b z_b - x_a x_2} \frac{\alpha}{3\pi M^2} \sqrt{1 - \frac{4m_l^2}{M^2}} \left(1 + \frac{2m_l^2}{M^2}\right) \frac{d\hat{\sigma}}{d\hat{t}}(a'b' \to \gamma^* X), \tag{59}$$

Here $x_a, z_a, x_b, z_b$ are related through specific kinematic expressions involving variables $x_1$, $x_2$ and $\tau$. The last part involves the differential cross section for the subprocess $a'b' \to \gamma^* X$ as a function of the momentum transfer squared $t^2$.

## Pomeron-Photon Associated Diffractive Production for Dilepton

Pomeron-photon associated diffractive production of dileptons is a hybrid process observed in high-energy physics, particularly within proton-proton collisions. This process combines the mechanisms of Pomeron exchange and photon exchange to produce a pair of leptons. It occurs when a Pomeron ($\mathbb{P}$), emanating from one of the protons, interacts with a photon emitted by the other proton, leading to the production of a dilepton pair shown in Figure 5, that explain the interaction of pomeron and photon leading to the production of dileptons $l^+l^-$.

Equation (60) represents a formula for the differential cross section of the Photon-Pomeron diffractive production of dileptons in a high-energy collision process. The expression includes a factorized product of PDFs, parton momenta fractions, and an alpha ($\alpha$) factor. The last part involves the differential cross section for the subprocess $a'b' \to \gamma^* X$ as a function of the momentum transfer squared, $t^2$. The entire expression represents the probability of observing the process $AB \to l^+l^- X$ through photon-Pomeron diffractive production, considering the various kinematic and partonic factors involved. It involves an integral over variables $x_a, x_b, z_b$, along with various parton distribution functions (PDFs) and form factors.

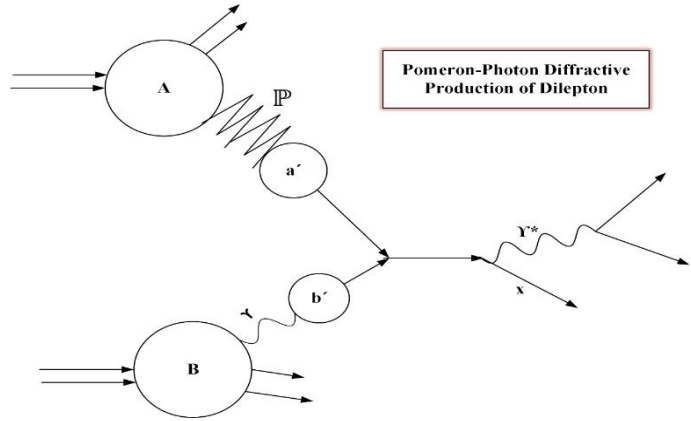

**Figure 5** Pomeron-Photon Diffractive Production of Dilepton ($l^+l^-$).

$f_{\mathbb{P}/A}(x_a, Q^2), f_{a'/\mathbb{P}}(z_a, \mu_F^2)$, are the parton distribution functions for Pomeron ($\mathbb{P}$) in particles $A$. $f_{\gamma/B}(x_b, Q^2), f_{b'/\gamma}(z_b, Q^2)$ are the parton distribution functions for photons in particles $B$.

$$\frac{d\sigma_{AB \to l^+l^- X}^{Pho.Pom,Diff}}{dM^2 dp_T^2 dy} = \int dx_a dx_b dz_b f_{\underset{A}{\mathbb{P}}}(x_a, Q^2) f_{\underset{\mathbb{P}}{a'}}(z_a, \mu_F^2) f_{\underset{B}{\gamma}}(x_b, Q^2) f_{\underset{\gamma}{b'}}(z_b, Q^2) \times \qquad (60)$$

$\frac{x_a z_a x_b z_b}{x_a x_b z_b - x_a x_2} \frac{\alpha}{3\pi M^2} \sqrt{1 - \frac{4m_l^2}{M^2}} \left(1 + \frac{2m_l^2}{M^2}\right) \frac{d\hat{\sigma}}{d\hat{t}}(a'b' \to \gamma^* X),$

Here $x_a, z_a, x_b, z_b$ are related through specific kinematic expressions involving variables $x_1, x_2$ and $\tau$. These relations describe the momentum fractions in terms of the partonic subprocesses and the squared invariant mass $M^2$.

$$x_a = \frac{x_b x_1 z_b - \tau}{x_b z_a z_b - x_2 z_a}, \qquad (61)$$

$$x_b = \frac{x_a z_a x_2 - \tau}{x_a z_a z_b - x_2 z_b}, \qquad (62)$$

$$z_a = \frac{x_b x_1 z_b - \tau}{x_b x_a z_b - x_2 x_a}, \qquad (63)$$

$$z_b = \frac{x_b x_1 z_a - \tau}{x_b x_a z_a - x_2 x_b}. \tag{64}$$

Putting the values of $x_a, z_a, x_b, z_b$, from equation to (61), (62), (63), (64), and PDF of Pomeron ($\mathbb{P}$) and Photon from equation (45) and (48) we will get the main equation for Pomeron-photon diffractive production of Dilepton.

## Diffractive Production for Photon

In order to move forward to describe the diffractive production for photon. Here, is a layout as shown in the Table 3 that defines the further direction of work regarding equations. It include the diffractive production of photon mainly in five steps. These are single pomeron and double pomeron diffractive production of photon, single photon and double photon production of photon, and on the last pomeron-photon diffractive production of photon.

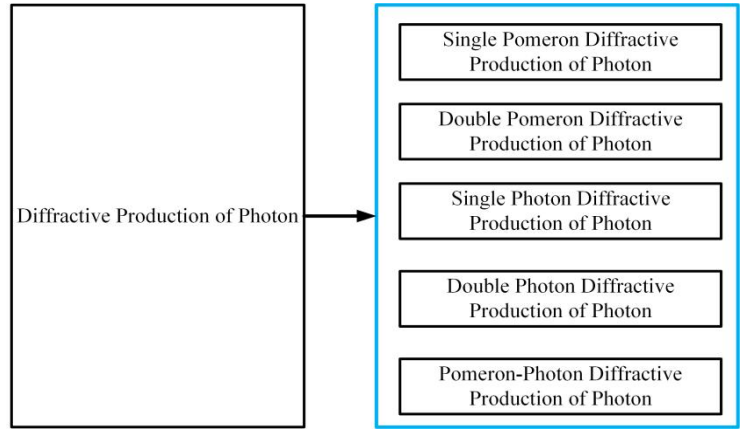

**Table 3** Layout for Diffractive Production of Photon

## Single-Pomeron Diffractive Production for Photon

Single Pomeron diffractive production processes for photons involve specific interactions in high-energy particle physics where a Pomeron ($\mathbb{P}$), a hypothetical particle associated with strong interactions, is exchanged. In these processes, only one of the colliding particles undergoes a diffractive interaction, resulting in the production of photons.

In order to get the main equation for the diffractive production of photon, Mandelstam variables, kinematics variables play an important role. The Mandelstam variables $\hat{s}$, $\hat{t}$ and $\hat{u}$ contains the equation (1), (2), and (3). These variables describe the square of the center-of-mass energy in the partonic subprocess $a'b \to \gamma X$ and square of the momentum transfer in the partonic subprocess. The variables $x_1, x_2$ are the momentum fractions of the partons in the pomeron involved in the collision shows in (27) and (28). Using the values of $x_1, x_2$, we will get the values of variables $x_a, x_b$ and $z_a$.

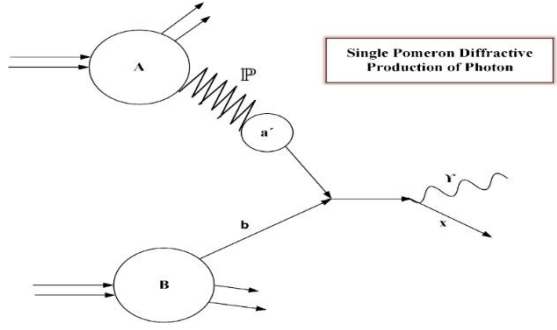

**Figure 6** Single Pomeron Diffractive Production of Photon ($\gamma$).

$$x_a = \frac{x_b x_1}{x_b z_a - x_2 z_a}, \tag{65}$$

$$x_b = \frac{x_a z_a x_2}{x_a z_a - x_1}, \tag{66}$$

$$z_a = \frac{x_b x_1}{x_b x_a - x_2 x_a}. \tag{67}$$

If we are going to describe the cross-section $d\hat{\sigma}/d\hat{t}$ for the production of photon we have, Equation (21) that represents the differential cross-section for the process where a quark-antiquark pair $q\bar{q}$ produces a photon ($\gamma$) and a gluon ($g$). Equation (22) represents the differential cross-section for the process where a quark-antiquark pair $q\bar{q}$ produces two photons ($\gamma$). Equation (23) represents the differential cross-section for the process where a quark and a gluon pair $qg$ produces a photon ($\gamma$) and a quark $q$. The structure of the equation is similar to the equation (21), but with different ratios of the Mandelstam variables.

The expression (68), represents the differential cross-section $d\sigma$ for single Pomeron diffractive production processes involving the production of photons $AB \rightarrow \gamma X$ in high-energy particle physics. The process involves colliding particles $A$ and $B$, with $A$ undergoing the diffractive interaction. Various kinematic variables are introduced, such as $p_T$ (transverse momentum), $y$ (rapidity), $x_a, z_a, x_b$ mathematically described in equations (65), (66), and (67). Parton distribution functions (PDFs) and factorization scale for the Pomeron and the parton $a'$ inside the Pomeron are represented respective.

$$\frac{d\sigma^{Sin.Pom.Diff}_{AB \rightarrow \gamma X}}{dp_T^2 dy} = \int dx_a \, dx_b f_{\mathbb{P}/A}(x_a, Q^2) f_{a'/\mathbb{P}}(z_a, \mu_F^2) f_{b/B}(x_b, Q^2) \frac{x_a z_a x_b}{x_a x_b - x_a x_2} \frac{d\hat{\sigma}}{d\hat{t}}(a'b \rightarrow \gamma X), \tag{68}$$

$f_{\mathbb{P}/A}(x_a, Q^2), f_{a'/\mathbb{P}}(z_a, \mu_F^2)$, represented in equation (9), (14), (15), and (16). These expressions describe the probability of observing single Pomeron diffractive production of photons in a high-energy collision, incorporating various factors such as parton distribution functions, factorization scales, and cross-sections for partonic subprocesses. The additional equations define relationships between kinematic variables involved in the process.

## Double-Diffractive Production (Pomeron-Pomeron) for Photon

Double diffractive production involving Pomeron-Pomeron interactions for photon production is a more complex process than single Pomeron exchange. In this scenario, two

Pomerons, each originating from the colliding particles (for example, in a proton-proton (pp) collision), interact with each other to produce a photon shown in Figure 7.

The main equation for double diffractive production of photon is:

$$\frac{d\sigma_{AB \to l^+l^-X}^{Dou.Pom.Diff}}{dp_T^2 dy} = \int dx_a\, dx_b\, dz_b f_{\frac{\mathbb{P}}{A}}(x_a, Q^2) f_{\frac{a'}{\mathbb{P}}}(z_a, \mu_F^2) f_{\frac{\mathbb{P}}{B}}(x_b, \mu_F^2) f_{\frac{b'}{\mathbb{P}}}(z_b, \mu_F^2) \qquad (69)$$

$$\times \frac{x_a z_a x_b z_b}{x_a x_b z_b - x_a x_2} \frac{d\hat{\sigma}}{d\hat{t}}(a'b' \to \gamma X),$$

The process involves colliding particles $A$ and $B$, with $A$ undergoing the diffractive interaction. Various kinematic variables are introduced, such as $p_T$ (transverse momentum), $y$ (rapidity), $x_a, z_a, x_b, z_b$ mathematically described in equation (70), (71), (72) and (73). $f_{\mathbb{P}/A}(x_a, Q^2), f_{a'/\mathbb{P}}(z_a, \mu_F^2), f_{\mathbb{P}/B}(x_a, Q^2), f_{b'/\mathbb{P}}(z_a, \mu_F^2)$ represented in equation (9), (14), (15), and (16). By putting all the equations, we can get our desire equation.

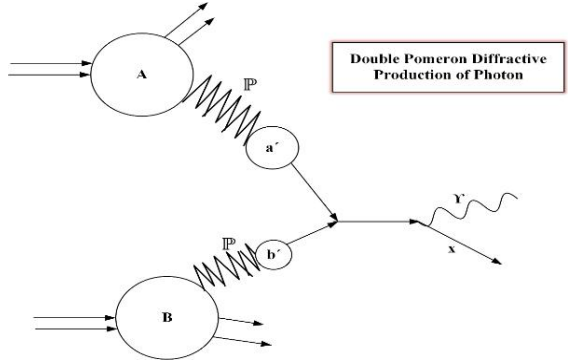

**Figure 7** Double Pomeron Diffractive Production of Photon ($\gamma$).

By using the values of Mandelstam variables from equation (1), (2), and (3), the values of momentum transfer $x_1, x_2$, from equation (27) and (28) we can get the values of variables $x_a, x_b$ and $z_a, z_b$.

$$x_a = \frac{x_b x_1 z_b}{x_b z_a z_b - x_2 z_a}, \qquad (70)$$

$$x_b = \frac{x_a z_a x_2}{x_a z_a z_b - x_2 z_b}, \qquad (71)$$

$$z_a = \frac{x_b x_1 z_b}{x_b x_a z_b - x_2 x_a}, \qquad (72)$$

$$z_b = \frac{x_a x_1 z_a}{x_b x_a z_a - x_2 x_b}. \qquad (73)$$

Use the cross-section $d\hat{\sigma}/d\hat{t}$ that are described in equation (21), equation (22) and equation (23) for the production of photon. Parton distribution functions (PDFs) and factorization scale for the Pomeron and the parton $a'$ inside the Pomeron are represented respectively.

## Single-Photon Diffractive Production for Photon

In single photon diffractive production, where hadron $A$ emits a photon that interacts diffractively with hadron B, resulting in the emission of another photon while both hadrons remain intact as diffracted remnants, $a'$ and $b$, this leads to the production of photon shown in Figure 8.

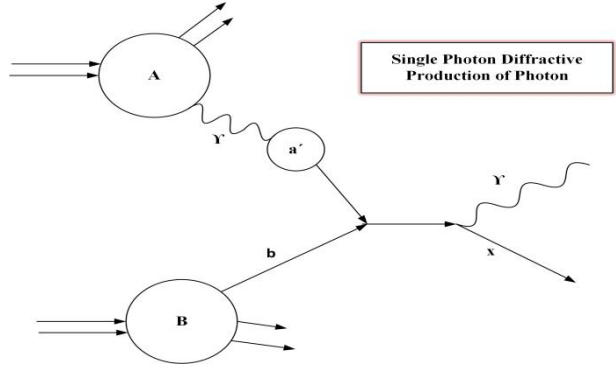

**Figure 8** Single Photon Diffractive production of Photon ($\gamma$).

This process is a quantum electro-dynamical event showcasing the exchange of a photon in a high-energy physics scenario. The main equation is represented by (74). In the main equation $d\sigma_{AB \to \gamma X}^{Sin.Pho.Diff}/dp_T^2 dy$, represents the differential cross-section for producing a photon in a single-photon diffractive process, differentiated by the transverse momentum squared $p_T^2$ and rapidity y of the photon. The integral over $dx_a dx_b$ incorporates the convolution of parton distribution functions (PDFs) and photon fluxes from the colliding particles A and B, integrating over the momentum fractions $x_a$, $x_b$ that the partons or photons carry from their respective parent particles.

$$\frac{d\sigma_{AB \to \gamma X}^{Sin.Pho.Diff}}{dp_T^2 dy} = \int dx_a \, dx_b f_{\frac{\gamma}{A}}(x_a, Q^2) f_{\frac{a'}{\gamma}}(z_a, \mu_F^2) f_{\frac{b}{B}}(x_b, Q^2) \frac{x_a z_a x_b}{x_a x_b - x_a x_2} \frac{d\hat{\sigma}}{dM^2 d\hat{t}}(a'b \to \gamma X), \quad (74)$$

$f_{\gamma/A}(x_a, Q^2)$ is the photon flux or distribution function in particle A, describing the probability of finding a photon with a fraction $x_a$ of particle, momentum at a scale $Q^2$. $f_{\gamma/B}(x_b, Q^2)$ is a PDF of particle B similar to $f_{\gamma/A}(x_a, Q^2)$. $f_{a'/\gamma}(z_a, \mu_F^2)$ represents the distribution function of a parton $a'$ (like a quark or gluon) within the photon, describing the probability of finding parton $a'$ with momentum fraction $z_a$, at a factorization scale $\mu_F^2$. PDF equations are given in equation (45) and (48). We have:

$$x_a = \frac{x_b x_1}{x_b z_a - x_2 z_a}, \quad (75)$$

$$x_b = \frac{x_a z_a x_2}{x_a z_a - x_1}, \quad (76)$$

$$z_a = \frac{x_b x_1}{x_b x_a - x_2 x_a}, \quad (77)$$

By using the values of Mandelstam variables from equation(1), (2), and (3), the values of momentum transfer $x_1, x_2$, from equation (27) and (28) we can get the values of variables $x_a, x_b$ and $z_a$.

## Double-Photon (Photon-Photon) Diffractive Production for Dilepton

Double-photon diffractive production of dileptons in high-energy collisions, like proton-proton collisions at the Large Hadron Collider (LHC), is a fascinating process that explores quantum electrodynamics (QED) interactions within a Quantum Chromodynamics (QCD) environment. In this process, that is shown in Figure 9, two photons from the protons involved in the collision interact to produce a pair of leptons, which are considered diffractive because they exchange virtual photons without leaving the colliding protons largely intact. The theoretical framework for this process includes the equivalent photon approximation (EPA), which models the photon flux from protons and calculates the effective number of photons surrounding a proton based on its electromagnetic field.

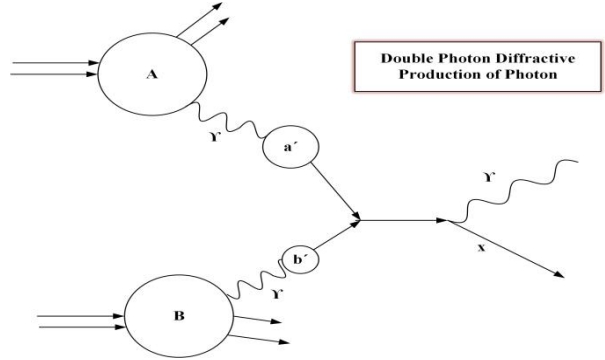

**Figure 9** Double Photon Diffractive Production of Photon ($\gamma$).

The cross-section for the photon-photon interaction leading to dilepton production is calculated within the framework of QED, and the diffractive nature of the event is characterized by intact protons after the collision and the presence of rapidity gaps, which indicate a non-destructive exchange mechanism. The main equation is given in equation (78). In the main equation $d\sigma_{AB \to \gamma X}^{Dou.Pho.Diff}/dp_T^2 dy$, represents the differential cross-section for producing a photon in a double-photon diffractive process, differentiated by the transverse momentum squared $p_T^2$ and rapidity $y$ of the photon. The integral over $dx_a dx_b$ incorporates the convolution of parton distribution functions (PDFs) and photon fluxes from the colliding particles $A$ and $B$, integrating over the momentum fractions $x_a, x_b$ that the partons or photons carry from their respective parent particles.

$$\frac{d\sigma_{AB \to l^+l^-X}^{Dou.Pho.Diff}}{dp_T^2 dy} = \int dx_a\, dx_b dz_b f_{\gamma/A}(x_a, Q^2) f_{a'/\gamma}(z_a, Q^2) f_{\gamma/B}(x_b, Q^2) f_{b'/\gamma}(z_b, Q^2) \times \frac{x_a z_a x_b z_b}{x_a x_b z_b - x_a x_2} \frac{d\hat{\sigma}}{d\hat{t}}(a'b' \to \gamma X), \tag{78}$$

$f_{\gamma/A}(x_a, Q^2), f_{\gamma/B}(x_b, Q^2)$ is the photon flux or distribution function in particle $A$ and $B$, describing the probability of finding a photon with a fraction $x_a, x_b$ of particle, momentum at a scale $Q^2$. $f_{a'/\gamma}(z_a, \mu_F^2), f_{b'/\gamma}(z_b, \mu_F^2)$ represents the distribution function of a parton $a', b'$ (like a quark or gluon) within the photon, describing the probability of finding parton $a', b'$ with momentum fraction $z_a, z_b$, at a factorization scale $\mu_F^2$. PDF equations are given in equation (45) and (48). We have:

$$x_a = \frac{x_b x_1 z_b}{x_b z_a z_b - x_2 z_a}, \tag{79}$$

$$x_b = \frac{x_a z_a x_2}{x_a z_a z_b - x_2 z_b}, \tag{80}$$

$$z_a = \frac{x_b x_1 z_b}{x_b x_a z_b - x_2 x_a}, \tag{81}$$

$$z_b = \frac{x_a x_1 z_a}{x_b x_a z_a - x_2 x_b}, \tag{82}$$

By using the values of Mandelstam variables from equation (1), (2), and (3), the values of momentum transfer $x_1, x_2$, from equation (27) and (28) we can get the values of variables $x_a, x_b$ and $z_a, z_b$. Use the cross-section $d\hat{\sigma}/d\hat{t}$ that are described in equation (21), equation (22) and equation (23) for the production of photon.

## Pomeron-Photon associated Diffractive Production for Photon

Pomeron-Photon associated Diffractive Production for a photon is a process where a neutral Pomeron ($\mathbb{P}$) from one hadron and a photon ($\gamma$) from another interact, producing a final-state photon, with the hadrons remaining mostly intact. In the Feynman diagram of this process, two incoming hadrons approaching each other, one hadron emitting a Pomeron (represented typically by a curly line with no arrow), the other hadron emitting a photon (represented by a wavy line), a vertex where the Pomeron and the photon interact. The emission of a final-state photon from this vertex, both hadrons continuing on their paths as remnants of the interaction shown in Figure 10.

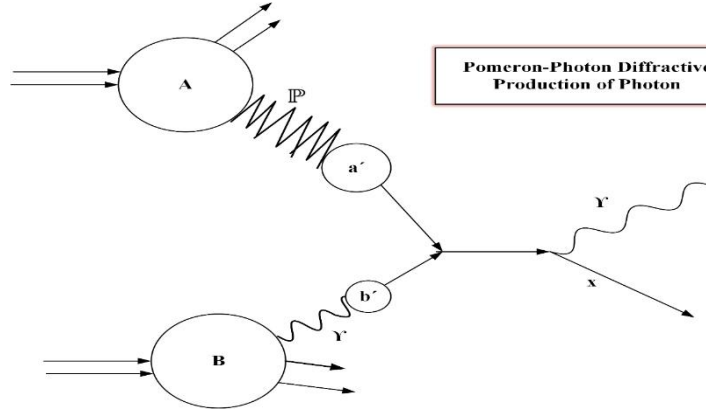

**Figure 10** Pomeron-Photon Diffractive Production of Photon ($\gamma$).

Each line and vertex in the diagram correlate with mathematical expressions that describe the probability of the interaction occurring and the energy and momentum of the particles involved.

In the main equation $d\sigma_{AB\to\gamma X}^{Pho.Pom.Diff}/dp_T^2 dy$, it represents the differential cross-section for producing a photon in a pomeron-photon diffractive process, differentiated by the transverse momentum squared $p_T^2$ and rapidity $y$ of the photon. The integral over $dx_a dx_b$ incorporates the convolution of parton distribution functions (PDFs) and photon fluxes from the colliding particles $A$ and $B$, integrating over the momentum fractions $x_a, x_b$ that the partons or photons carry from their respective parent particles. $f_{\mathbb{P}/A}(x_a, Q^2)$, is a distribution function of the Pomeron within particle $A$, showing the probability density of finding a Pomeron carrying a momentum fraction, describing the probability of finding a photon with a fraction $x_a$, of particle, momentum at a scale $Q^2$.

$$\frac{d\sigma_{AB\to l^+l^-X}^{Pho.Pom,Diff}}{dp_T^2 dy} = \int dx_a\, dx_b dz_b f_{\underset{A}{\mathbb{P}}}(x_a, Q^2) f_{\underset{\mathbb{P}}{a'}}(z_a, \mu_F^2) f_{\underset{B}{\gamma}}(x_b, Q^2) f_{\underset{\gamma}{b'}}(z_b, Q^2) \times \frac{x_a z_a x_b z_b}{x_a x_b z_b - x_a x_2} \frac{d\hat{\sigma}}{d\hat{t}}(a'b' \to \gamma X), \tag{83}$$

$f_{\gamma/B}(x_b, Q^2)$ is the photon flux or distribution function in particle $B$ and $f_{b'/\gamma}(z_b, \mu_F^2)$ is the

distribution of parton $b'$ within this photon, respectively. $f_{a'/\mathbb{P}}(z_a, \mu_F^2)$, describing the probability of finding parton $a'$ with momentum fraction $z_a$, at a factorization scale $\mu_F^2$. PDF equations are given in equation (45) and (48). We have:

$$x_a = \frac{x_b x_1 z_b}{x_b z_a z_b - x_2 z_a}, \tag{84}$$

$$x_b = \frac{x_a z_a x_2}{x_a z_a z_b - x_2 z_b}, \tag{85}$$

$$z_a = \frac{x_b x_1 z_b}{x_b x_a z_b - x_2 x_a}, \tag{86}$$

$$z_b = \frac{x_a x_1 z_a}{x_b x_a z_a - x_2 x_b}, \tag{87}$$

By using the values of Mandelstam variables from equation (1), (2), and (3), (2), and (3), the values of momentum transfer $x_1, x_2$, from equation (27) and (28) we can get the values of variables $x_a, x_b$ and $z_a, z_b$. Use the cross-section $d\hat{\sigma}/d\hat{t}$ that are described in equation (21), equation (22) and equation (23) for the production of photon. By using all the terms, we can fine the main equation of pomeron-photon diffractive production of photon.

## Parton Distribution Function (PDF) of Proton and Neutron

The equations (88) and (91), presented are expressions for the structure function $F_2^{ep}$ of the proton, denoted as $F_2^{en}(x_B)$ and the neutron, denoted as $F_2^{en}(x_B)$, which are measured in deep inelastic scattering experiments. The structure function $F_2$ gives information about the distribution of quarks within a nucleon (proton or neutron) as a function of the Bjorken scaling variable $x_B$, which represents the fraction of the nucleon's momentum carried by the parton (quark or gluon). In these expressions, $e_i$ is the charge of the quark of flavor $i$, in units of the elementary charge. $x_B f_i(x_B)$ represents the Parton Distribution Function (PDF) for a quark of flavor $i$, which gives the probability density for finding a quark with momentum fraction $x_B$ inside the nucleon. The sums are over all quark flavors $i$, where $u, d, c, s, b$, represent the up, down, charm, strange, and bottom quarks, respectively. The bar denotes the corresponding antiquarks.

For the proton, $F_2^{ep}(x_B)$ includes a sum of the PDFs weighted by the square of the quark charges, with the factor 4/9 for the up and charm quarks (having a charge of +2/3 in units of elementary charge), and 1/9 for the down, strange, and bottom quarks (having a charge of −1/3). For the neutron $F_2^{en}(x_B)$ is similar, but the roles of the up and down quarks are switched due to the neutron's different quark composition.

$$F_2^{ep}(x_B) = \sum_i e^2{}_i x_B \, f_i(x_B), \tag{88}$$

$$F_2^{ep}(x_B) = \frac{4}{9} x_B [u(x_B) + \bar{u}(x_B) + c(x_B) + \bar{c}(x_B)], \tag{89}$$

$$F_2^{ep}(x_B) = \frac{1}{9} x_B [d(x_B) + \bar{d}(x_B) + s(x_B) + \bar{s}(x_B) + b(x_B) + \bar{b}(x_B)], \tag{90}$$

$$F_2^{en}(x_B) = \sum_i e^2{}_i x_B \, f_i(x_B), \qquad (91)$$

$$F_2^{en}(x_B) = \frac{4}{9} x_B [d(x_B) + \bar{d}(x_B) + c(x_B) + \bar{c}(x_B)], \qquad (92)$$

$$F_2^{en}(x_B) = \frac{1}{9} x_B [u(x_B) + \bar{u}(x_B) + s(x_B) + \bar{s}(x_B) + b(x_B) + \bar{b}(x_B)], \qquad (93)$$

Exploring Parton Distribution Functions (PDFs) for protons and neutrons, alongside their structure functions, is pivotal for advancing our understanding of the intricate structure of nucleons and the dynamics governing subatomic particles. It provides an in-depth look into how quarks are distributed within protons and neutrons, revealing the complexities of the strong force interactions that glue these fundamental constituents together.

## III. Numerical Results

This section will feature graphical representations to illustrate the dependence of the cross-section on variables such as the invariant mass of the dilepton system and momentum fractions, alongside the kinematic distributions like rapidity and transverse momentum of the produced dileptons. Focusing on various collision configurations differentiated by atomic numbers, charges, and center-of-mass energies. This analysis spans collisions ranging from heavy ion pairs like Au-Au and Pb-Pb to proton-induced interactions with heavy ions and other protons, across energy scales from 200 GeV at RHIC to 13 TeV at LHC as mentioned in Table 4. These configurations provide a comprehensive dataset for exploring diffractive production of dileptons and photons, revealing the significant role of nuclear size, charge, and collision energy.

Table 4 LHC Collision Parameters for Diffractive Production Studies

| | | | |
|---|---|---|---|
| **Pb-Pb** | A=208 | Z=82 | $\sqrt{s_{NN}}$=5.02TeV |
| **P-Pb** | A=208 | Z=82 | $\sqrt{s_{NN}}$=8.16TeV |
| **P-p** | A=1 | Z=1 | $\sqrt{s_{NN}}$=7TeV & 13TeV |

By using these different energies and parameters we will get the results, represents the compendium of collision configurations and their corresponding center-of-mass energies utilized in diffractive production studies Large Hadron Collider (LHC), respectively. With lead-lead (Pb-Pb), proton-lead (p-Pb), and proton-proton (p-p) collisions at TeV energy scales, the LHC allows for the detailed exploration of diffractive processes that are sensitive to the energy density and the structure of the colliding bodies. Pb-Pb collisions, at an energy of 5.02 TeV, these collisions provide a vast landscape to study high-energy heavy-ion interactions and the resulting high-temperature QCD phenomena. p-Pb collisions, with a center-of-mass energy of 8.16 TeV, these asymmetric collisions are crucial for understanding the interaction between a single proton and the dense nuclear field of the lead ion. p-p collisions, occurring at 7 and 13 TeV.

The differential cross-sections, as shown in Figure 11, provide significant insights into the energy-dependent behavior of diffractive processes. By using the derived equation (20) of single pomeron diffractive production, equation (39) of double pomeron diffractive production, equation (53) and (59) for single photon and double photon diffractive production for dilepton and lastly the pomeron-photon's equation (60) for the production of dilepton with using all the variables we get the production of dilepton at LHC. Figure 11 explains the invariant cross section of large- $p_T$ dilepton production for pb-pb collisions at energy $\sqrt{s_{NN}}$ = 5.0 TeV, shown in top left. The dashed line (blue) is for the single pomeron process, the dotted line (brown) is for the double pomeron diffractive processes, the dotted line (red) is for the single photon production processes, the dashed- dotted line (sea-foam green) is for double photon processes, the dashed-dotted line (olive green) is for the pomeron-photon production for dilepton. While other three results are same as explained but for different collisions. Top right is for p-Pb collisions at $\sqrt{s_{NN}}$= 8.16 TeV. Bottom left is for p-p collisions at $\sqrt{s_{NN}}$= 7 TeV. Bottom right is for p-p collisions at $\sqrt{s_{NN}}$= 13 TeV. With the same invariant mass of 600MeV<M<900MeV.

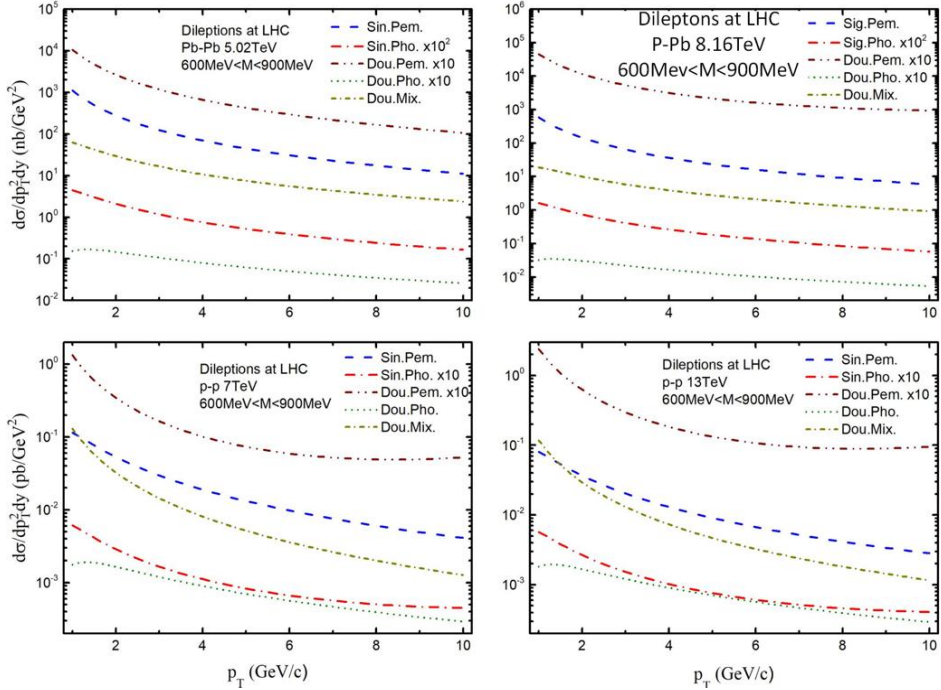

Figure 11 The cross sections for e$^+$e$^-$ production from the single and double diffractive processes at LHC.

The cross-sections diminish with an increase in transverse momentum ($p_T$), which is a characteristic trend expected from diffractive scattering phenomena. The results display differential cross sections as a function of transverse momentum ($p_T$) for diffractive dilepton production in various collision environments at the LHC. The y-axis shows the differential cross-section ($d\sigma/dp_T^2 dy$) which measures how often particles are produced at a certain transverse momentum ($p_T$, shown on the x-axis) and rapidity (*y*). The $p_T$ is given in units of GeV/c, indicating the momentum perpendicular to the beam direction. The lines in each graph represent different production processes:

- Sin.Pem.: Single diffractive production with a single pomeron exchange. The solid lines (blue) indicate this process and it's typically the baseline for comparison.
- Sin.Pho.: Single diffractive production with photon exchange. This process is depicted with dotted lines (red) and scaled by a factor for visibility (e.g., x10), given that photon exchange processes are less probable than pomeron exchanges.
- Dou.Pem.: Double diffractive production with double pomeron exchange. The dashed lines (brown) show this process, which is generally rarer than single diffractive processes and therefore often scaled up by a significant factor (e.g., x10) to be visible on the same scale.

- ➢ Dou.Pho.: Double diffractive production involving photons. The dot-dashed lines (sea-foam green) represent this and are again scaled for visibility due to their lower cross-sections.
- ➢ Dou.Mix.: This might represent a mixed double exchange involving both pomeron and photon exchanges. The fine dotted lines (olive green) depict this process.

**Cross-Section Hierarchies in LHC Collision Events for Dilepton**

- Pb-Pb 5.02 TeV

The hierarchy of processes by the size of their cross-sections is as follows:

Single Photon (Sin Pho) production has the lowest cross-section. Double Photon (Dou Pho) production is higher than single photon production. Mixed Double (Dou Mix) exchange is greater than double photon production. Single Pomeron (Sin Pem) and Double Pomeron (Dou Pem) are approximately equal and have the highest cross-sections among the processes considered.

- p-Pb 8.16 TeV

The ordering is similar to Pb-Pb collisions, with the distinction that Double Pomeron (Dou Pem) production has a larger cross-section than Single Pomeron (Sin Pem) production.

- p-p 7 TeV

The hierarchy is much the same as in Pb-Pb collisions, with Single Pomeron and Double Pomeron processes being nearly equal and the largest.

- p-p 13 TeV

Again, Single Photon and Double Photon production have the lowest cross-sections, but here the Double Pomeron process becomes more dominant compared to the Single Pomeron process.

## Comparative Observation

When moving from Pb-Pb to p-Pb collisions, the Double Pomeron process increases in significance, suggesting that Double Pomeron interactions play a more crucial role in lighter collision systems. The differential cross-sections for other processes remain nearly unchanged as the energy increases, indicating that their contributions to dilepton production do not significantly depend on the system size. As the energy increases from 7 TeV to 13 TeV in p-p collisions, the Single Pomeron cross-section decreases slightly, while the Double Pomeron cross-section increases, indicating that Double Pomeron processes become more significant at higher energies. Other processes' differential cross-sections do not show significant changes with the increase in energy. From these observations, Double Pomeron processes tend to become more prominent as we move to higher energies and lighter collision systems. The consistent trend across different energies and systems, where differential cross-sections decrease with increasing $p_T$, reflects the fundamental dynamics of particle production in high-energy collisions. Magnitude of cross sections for processes involving pomeron exchanges are higher than those with photon exchanges, reflecting the stronger coupling of the pomeron in diffractive processes. There is a visible difference in the cross sections when comparing 7 TeV and 13 TeV proton-proton collisions. By comparing the single and double pomeron exchanges, it is noticeable that the double exchange processes have a much lower cross-section. This is expected due to the lower probability of two independent pomeron exchanges occurring simultaneously.

## Analyzing Diffractive Photon Production in LHC Collisions

The graphical results presented in Figure 12 illustrate the differential cross-section measurements for photon production through diffractive processes at the Large Hadron Collider (LHC), across various collision energies and systems. The differential cross-sections, as shown in Figure 12 provide significant insights into the energy-dependent behavior of diffractive processes. By using the derived equation (68) of single pomeron diffractive production, equation (69) of double pomeron diffractive production, equation (74) and (78) for single photon and double photon diffractive production for photon and lastly the pomeron-photon's equation (83) for the production of photon with using all the variables we get the production of dilepton at LHC. In order to analyze the graphical results, each plot displays the differential cross-section ($d\sigma/dp_T^2 dy$) as a function of the transverse momentum ($p_T$) of the photons produced in the collisions. Figure 12 explains the invariant cross section of large- $p_T$ photon production for pb-pb collisions at energy $\sqrt{s_{NN}}$ = 5.0 TeV, shown in top left. The dashed line (blue) is for the single pomeron process, the dotted line (brown) is for the double pomeron diffractive processes, the dotted line (red) is for the single photon production processes, the dashed- dotted line (sea-foam green) is for double photon processes, the dashed-dotted line (olive green) is for the pomeron-photon production for photon.

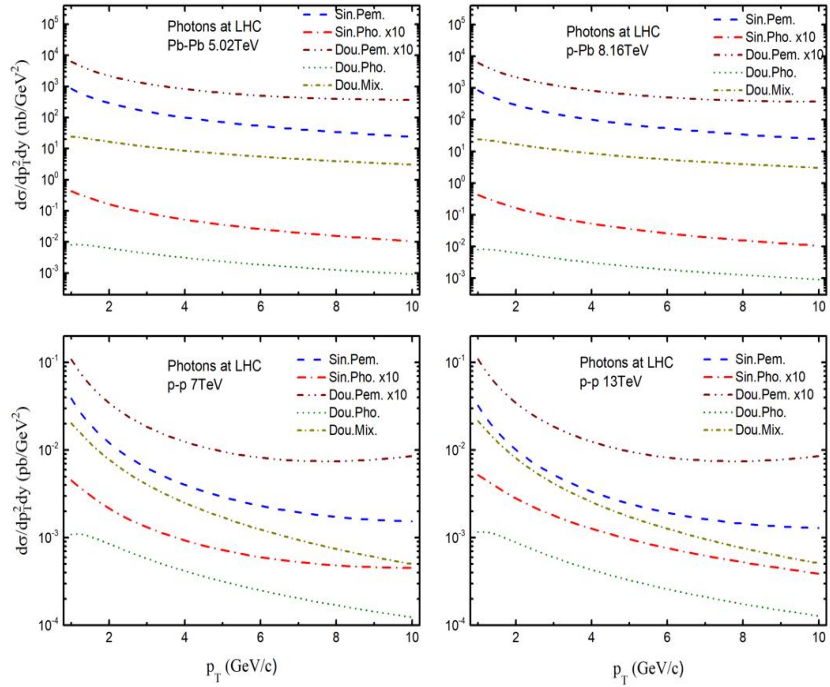

**Figure 12** The cross sections for photon production from the single and double diffractive processes at LHC.

While other three results are same as explained but for different collisions. Top right is for p-Pb collisions at $\sqrt{s_{NN}}$= 8.16 TeV. Bottom left is for p-p collisions at $\sqrt{s_{NN}}$= 7 TeV. Bottom right is for p-p collisions at $\sqrt{s_{NN}}$= 13 TeV. The lines within each plot denote:

- Sin.Pem. (Single Pomeron Exchange):
  The dashed blue lines likely represent the photon production via single pomeron exchange, an important process in understanding the soft interactions at play in such high-energy collisions.
- Sin.Pho. (Single Photon Exchange):
  The dot-dashed red lines, adjusted with a multiplicative factor (x10), indicate photon production through a single photon exchange, a rarer process compared to pomeron exchanges due to the electromagnetic nature of photon interactions.
- Dou.Pem. (Double Pomeron Exchange):
  The dotted brown lines, enhanced for visibility with a factor (x10), depict the even less

- probable double pomeron exchange events.
  - ➢ Dou.Pho. (Double Photon Exchange):
    The fine sea-foam green dotted lines suggest an even rarer occurrence of photon production via double photon exchange.
  - ➢ Dou.Mix. (Mixed Double Exchange):
    The long dash-dot olive green pattern could illustrate a combination of pomeron and photon exchanges in a double diffractive process.

**Cross-Section Hierarchies in LHC Collision Events for Photon**

- Pb-Pb Collisions at 5.02 TeV

The single photon production process (Sin. Pho.) has the lowest cross-section. Double photon production (Dou. Pho.) has a higher cross-section than single photon production. The mixed double exchange process (Dou. Mix.) has a higher cross-section than double photon production. The single pomeron exchange (Sin. Pem.) and double pomeron exchange (Dou. Pem.) have similar cross-sections, which are the highest among the processes displayed.

- p-Pb Collisions at 8.16 TeV

Single photon production (Sin. Pho.) remains the lowest. Double photon production (Dou. Pho.) has a higher cross-section. The mixed double exchange (Dou. Mix.) has a higher cross-section than both photon processes. Single pomeron exchange (Sin. Pem.) has a slightly smaller cross-section than double pomeron exchange (Dou. Pem.), which is the highest.

- p-p Collisions at 7 TeV

Single pomeron exchange (Sin. Pem.) has a larger cross-section than double pomeron exchange (Dou. Pem.).

- p-p Collisions at 13 TeV

Single photon production (Sin. Pho.) continues to have the lowest cross-section. Double photon production (Dou. Pho.) has a higher cross-section than single photon production. The mixed double exchange (Dou. Mix.) has a higher cross-section than double photon production. Single pomeron exchange (Sin. Pem.) has a large cross-section than double pomeron exchange (Dou. Pem.).

## Comparative Observations

As the collision energy increases, the single pomeron exchange process tends to become less significant, especially in p-p collisions at 13 TeV. The cross-sections for the single pomeron exchange process (Sin. Pem.) appear to decrease relative to the double pomeron exchange process (Dou. Pem.) as the energy increases from 7 TeV to 13 TeV in p-p collisions.

## IV. Conclusion

The results suggests that in photon production at the LHC, as the energy of the collisions increases, the double pomeron exchange becomes more significant relative to single pomeron exchange. This trend is consistent across different types of collisions. The single and double photon production processes generally have lower cross-sections compared to pomeron-related processes. These findings are instrumental for understanding the underlying physics of photon production in high-energy collisions and may have implications for the study of Quantum Chromodynamics (QCD) and other aspects of particle physics.

Overall, the two results for dilepton and photon contain differential cross-section plots for particle production at the Large Hadron Collider (LHC) for different collision systems and energies. The first result deals with dilepton production, and the second with photon production. Here's a combined summary of the observations from both:

**Dilepton Production:**
- Cross-sections for all processes decrease with increasing $p_T$.
- For Pb-Pb at 5.02 TeV, the ordering from smallest to largest cross-section is: single photon, double photon, mixed double, single pomeron, and double pomeron.
- For p-Pb at 8.16 TeV and p-p at 13 TeV, the Double Pomeron has a larger cross-section compared to the Single Pomeron.
- For p-p at 7 TeV, Single Pomeron and Double Pomeron have similar cross-sections.
- Between Pb-Pb 5.02 TeV and p-Pb 8.16 TeV, the Double Pomeron becomes more pronounced.
- Between p-p 7 TeV and p-p 13 TeV, the Double Pomeron gains in significance as the energy increases.

**Photon Production:**
- As in dilepton production, photon production cross-sections also decrease with increasing $p_T$.
- Across all energies and collision systems, the general hierarchy is consistent: Single Photon < Double Photon < Mixed Double < Single Pomeron ≲ Double Pomeron.
- Double Pomeron exchange tends to increase in significance with increasing collision energy.

In order to explain the combined conclusion for both dilepton and photon production at the LHC, there is a consistent pattern across different types of collisions (Pb-Pb, p-Pb, and p-p) and varying energies (5.02 TeV, 7 TeV, 8.16 TeV, and 13 TeV):

All processes show decreasing cross-sections with increasing $p_T$, reflecting the fundamental behavior of such high-energy collisions where high $p_T$ events are less probable. Pomeron exchange processes (both Single and Double) generally have higher cross-sections compared to photon exchange processes. This suggests that strong interactions (possibly mediated by gluons in the case of pomerons) are dominant in these collision scenarios. The relative importance of Double Pomeron exchange grows with increasing energy, particularly in proton-proton collisions. This may indicate that as the energy available for the collisions increases, the probability of more complex interactions, like Double Pomeron exchanges, also increases. These observations are crucial for understanding the fundamental processes involved in particle collisions at high energies and contribute to the fields of Quantum Chromodynamics and particle physics research.

For dilepton production, at lower energies, such as in Pb-Pb collisions at 5.02 TeV, the cross-sections for single and double diffractive processes seem to be closer in magnitude across the $p_T$ spectrum. However, as the collision energy increases, as seen in p-Pb at 7 TeV and p-p at 13 TeV, there is a more pronounced difference between the single and double diffractive processes, with the single diffractive cross-sections typically being higher than those for double diffractive processes. The trend indicates that at higher energies, single diffractive processes become more dominant over double diffractive processes for dilepton production. This might be attributed to the higher available energy enhancing the single diffractive mechanism.

For photon production, a similar trend is observed in photon production. At lower energies (Pb-Pb at 5.02 TeV and at 8.16TeV), the single and double diffractive processes have cross-sections that are relatively close, but as we move to higher energies (p-p at 7 TeV and 13

TeV), the gap between the single and double diffractive processes widens, with single diffractive processes typically having a higher cross-section. At the highest energies shown (p-Pb at 8.16 TeV and p-p at 13 TeV), the single diffractive process is again observed to be the dominant mechanism for photon production.

Comparatively, each type of collision and energy configuration offers a different scale for the interaction, which affects the cross-section. Higher energies typically provide a higher probability for parton interactions, as indicated by the data at 13 TeV compared to 7 TeV for p-p collisions. The results for Pb-Pb and p-Pb collisions exhibit a difference in the absolute cross-sections due to the different nature of interactions in heavy-ion collisions compared to those involving protons. This comprehensive analysis is imperative for validating theoretical predictions about photon production in diffractive processes and serves as a cornerstone for future experimental planning and theoretical advancements in high-energy particle physics.

## V. ACKNOWLEDGEMENTS

This work is supported by Heilongjiang Science Foundation Project under grant No. LH2021A009, National Natural Science Foundation of China under grant No. 12063006, and Special Basic Cooperative Research Programs of Yunnan Provincial Undergraduate Universities Association under grant No. 202101BA070001-144. The National Natural Science Foundation of China under Grant No. 12165010, the Yunnan Province Applied Basic Research Project under grant No. 202101AT070145, the Xingdian Talent Support Project, the Young Top-notch Talent of Kunming, and the Program for Frontier Research Team of Kunming University 2023.